%% file: hadron2011.tex
\documentclass[a4paper,11pt]{article}

\usepackage{contribution}

\input{econfmacros}
\input{contributionmacros}

\begin{document}

\input{contribution}

\end{document}

%% file: econfmacros.tex
\newcommand{\weblink}[2][]{%
    \ifthenelse{\equal{#1}{}}%
    {\textnormal{\url{#2}}}%
    {\textnormal{\href{#2}{#1}}}%
}

\def\beq{\begin{equation}}
\def\eeq#1{\label{#1}\end{equation}}
\def\eeqn{\end{equation}}

\def\beqa{\begin{eqnarray}}
\def\eeqa#1{\label{#1}\end{eqnarray}}
\def\eeqan{\end{eqnarray}}

\let\bar=\overbar

\def\Dslash{\not{\hbox{\kern-4pt $D$}}}
\def\dslash{\not{\hbox{\kern-2pt $\del$}}}

\def\msb{{\bar{\ssstyle M \kern -1pt S}}}

%% file: contributionmacros.tex
\newcommand{\contribution}[7][]{%
  \clearpage
  \thispagestyle{plain}
  \ifthenelse{\equal{#1}{}}
  {\hypersetup{pdftitle={#2}}}
  {\hypersetup{pdftitle={#1}}}
  \hypersetup{pdfauthor={{#3} {#4}}}
  {\centering\normalfont\LARGE\bfseries\sffamily #2 \par\nobreak}
  \lhead{}
  \chead{%
    \textit{\footnotesize XIV International Conference on Hadron Spectroscopy
      (\weblink[\textit{hadron2011}]{http://www.hadron2011.de}), 13-17 June 2011, Munich, Germany}%
  }
  \rhead{}
  \bigskip
  \begin{center}
    {#3} {#4}\ifthenelse{\equal{#6}{}}{}{\footnote{\weblink[#6]{mailto:#6}}}
    \ifthenelse{\equal{#7}{}}{}{#7} \\
    \textit{#5}
  \end{center}
  \bigskip
}

\renewcommand{\abstract}[1]{%
  \begin{center}
    \begin{minipage}{0.85\textwidth}
      \begin{footnotesize}
        #1
      \end{footnotesize}
    \end{minipage}
  \end{center}
  \bigskip
}

%% file: contribution.tex
%
%
%
%
%
{  

\newcommand{\raa}{\ensuremath{R_{\rm{}AA}}}
\newcommand{\rcp}{\ensuremath{R_{\rm{}CP}}}
\newcommand{\jpsi}{J{\ensuremath{/\psi}}}
\newcommand{\jpsiee}{J\ensuremath{/\psi\rightarrow{}e^{+}e^{-}}}
\newcommand{\jpsimm}{J\ensuremath{/\psi\rightarrow{}\mu^{+}\mu^{-}}}
\newcommand{\epem}{\ensuremath{\rm{}e^{+}e^{-}}}
\newcommand{\mpmm}{\ensuremath{\mu^{+}\mu^{-}}}
\newcommand{\rsnn}{\ensuremath{\sqrt{s_{\rm{}NN}}}}
\newcommand{\rs}{\ensuremath{\sqrt{s}}}
\newcommand{\pt}{\mbox{$p_{\mathrm{t}}$}}
\newcommand{\qgp}{quark-gluon plasma}
%

\contribution[Quarkonia Measurements with ALICE at the LHC]  
{Quarkonia Measurements with ALICE at the LHC}  
{Frederick}{Kramer}  
{Institut f\"ur Kernphysik \\
  Johann Wolfgang Goethe--Universit\"at \\
  D 60438 Frankfurt, GERMANY}  
{kramer@ikf.uni-frankfurt.de}  
{on behalf of the ALICE Collaboration}  
%

\abstract{%
  ALICE is the dedicated heavy-ion experiment at the Large Hadron Collider (LHC). 
  It is designed to provide excellent capabilities to study the \qgp{} (QGP) in the highest energy density regime opened up by the LHC.
  Quarkonia are crucial probes of the QGP.
  High-precision data from pp collisions are an essential baseline, and serve as a crucial test for competing models of quarkonium hadroproduction.
  ALICE measures quarkonia down to $\pt{}=0$ via their decay channels into \epem{} at central ($|y| < 0.9$) and into \mpmm{} at forward rapidity ($-4.0 < y < -2.5$).
  We present first results on the transverse momentum and rapidity distributions of the inclusive \jpsi{} production cross section in $\rs{}=7$ and 2.76 TeV pp collisions.
  The dependence of the \jpsi{} yield on the charged particle multiplicity in $\rs{}=7$~TeV pp collisions is discussed.
  Finally, results on the inclusive \jpsi{} production in $\rsnn{}=2.76$~TeV Pb--Pb collisions, the nuclear modification factor \raa{} and the central-to-peripheral modification factor \rcp{} are shown.
}
%

Various models are describing the production of quarkonia in hadronic reactions such as proton-proton collisions.
Three of the most popular ones are the Color Singlet Model, the nonrelativistic QCD approach (NRQCD) and the Color Evaporation Model~\cite{Lansberg:2006dh, PhysRevD.23.1521, brambilla}.
So far their predictions do not fully reproduce the experimental observations such as differential cross sections or polarization, or many free parameters limit their predictive power.
Precise measurements at a new energy regime will give new constraints to theory.
Quarkonia, such as the \jpsi{} serve as a crucial probe to study the \qgp{} (QGP) in heavy-ion collisions.
The original prediction was a strong suppression of quarkonia yields~\cite{matsui-satz} by color-screening of the $Q\bar{Q}$ pairs in the colored medium.
This effect might depend on the binding radius of the different states and on the temperature of the fireball~\cite{satztemp}; thus, a stronger suppression of higher excited states could even reflect the temperature of the medium.
In central heavy-ion collisions at LHC energies \jpsi{} yields could also be affected in another way:
charm quark pairs are abundant enough so that a statistical or kinetic \jpsi{} formation out of uncorrelated $c\bar{c}$ pairs might become dominant~\cite{Andronic:2003zv, Thews:2005fs}.
Furthermore, initial state effects such as shadowing -- the modification of the parton distribution function of a nucleon inside a nucleus~\cite{brambilla} -- have to be taken into account.
Final state effects like nuclear absorption are expected to be negligible at LHC energies.
A measurement of such effects in p--A collisions will be crucial for drawing final conclusions; any modification of the yields stronger than expected will provide insights into the created medium.

The ALICE~\cite{alice:PPRv2} setup can be divided into two parts: the central rapidity detectors ($|\eta|<0.9$) and the forward muon spectrometer ($-4.0<\eta<-2.5$).
For this analysis, in the central barrel, mainly two detectors are used: the Inner Tracking System (ITS) and the Time Projection Chamber (TPC).
The large-volume TPC serves as the main tracking device.
Furthermore it provides particle identification via the specific energy loss d$E/$d$x$ in the detector gas.
Consisting of two layers each of silicon pixel, silicon strip and silicon drift detectors, the ITS provides high precision vertex reconstruction and tracking, improving the momentum resolution.
The muon spectrometer consists of a 10 plane cathode pad chamber tracking system behind a frontal absorber and a 4 plane resistive plate chamber trigger system further downstream behind an iron filter wall.
Muons with a momentum above 4~GeV/$c$ are filtered and detected.
In both channels, quarkonia are measured over a broad range of transverse momenta down to $p_{\rm t}=0$.
Finally, the VZERO detector, made of two scintillator arrays covering the pseudo rapidity ranges $-3.7<\eta{}<-1.7$ and $2.8<\eta{}<5.1$, is used for triggering purposes and collision centrality determination.
\begin{figure}[tb]
  \begin{center}
    \includegraphics[width=0.9\textwidth]{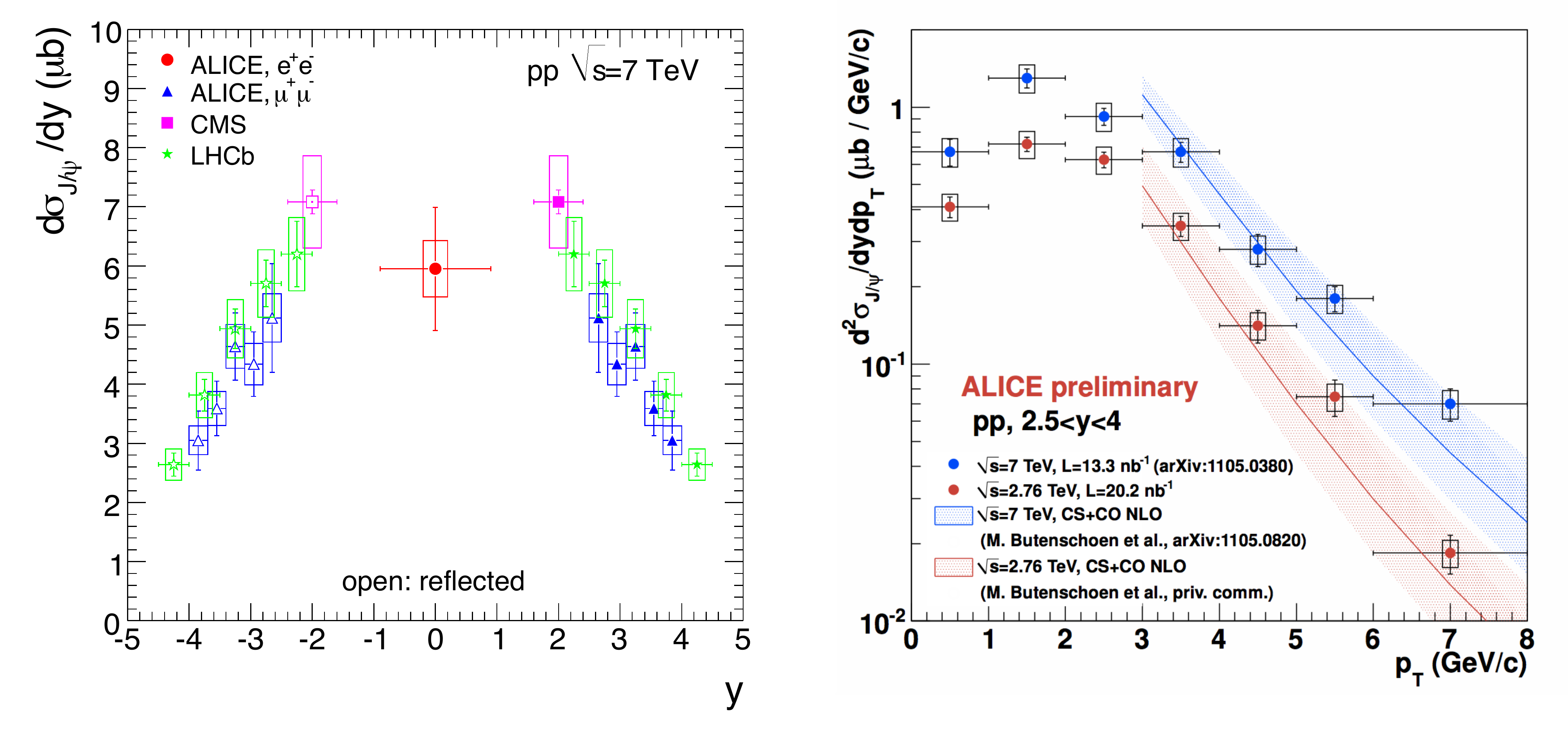}
    \caption{Differential \jpsi{} cross sections versus rapidity in 7 TeV pp collisions (left) and transverse momentum in 7 and 2.76 TeV pp collisions (right)~\cite{jppaper, butenschoen, lhcb, cms}}
    \label{fig:jp_pp}
  \end{center}
\end{figure}
%
\begin{table}[htb]
  \begin{center}
    \begin{tabular}{lcccc}  
      Energy    & System  & Year & L(\jpsiee{})      & L(\jpsimm{})\\
      \hline
      7 TeV     & pp      & 2010 & 3.9 nb$^{-1}$     & 15.6 nb$^{-1}$\\
      2.76 TeV  & pp      & 2011 & 1.1 nb$^{-1}$     & 20.2 nb$^{-1}$\\
      2.76 TeV  & Pb--Pb  & 2010 & 2.7 $\mu$b$^{-1}$ & 2.7 $\mu$b$^{-1}$\\
    \end{tabular}
    \caption{Integrated luminosities of the analyzed data samples}
    \label{tab:data}
  \end{center}
\end{table}
%

The \jpsi{} production was measured in $\rs{}=7$ and 2.76 TeV proton-proton collisions~\cite{jppaper, qmarnaldi}.
Table~\ref{tab:data} shows the integrated luminosities of the two samples, both for the data recorded in the central barrel and the muon spectrometer.
For the central barrel analysis, these are corresponding to the minimum bias trigger, defined as the logical OR between the requirement of at least one hit in the two ITS pixel layers, and a signal in one of the two VZERO detectors.
For the muon analysis, also a coincidence with a track reconstructed by the muon trigger electronics is required.
Differential cross sections have been obtained by the analysis of invariant mass spectra in different \pt{} and $y$ bins~\cite{jppaper}.
Presented are inclusive results, containing contributions from direct \jpsi{} production and feed-down from higher mass quarkonia and B hadron decays.
The left panel of Fig.~\ref{fig:jp_pp} shows the obtained rapidity dependence of the \jpsi{} production at $\rs{}=7$~TeV.
Red circles correspond to the measurement in the \epem{} ($|y|<0.9$), blue triangles to the result in the \mpmm{} ($-4.0<y<-2.5$) decay channels. 
While at forward rapidity results of other experiments (green: LHCb~\cite{lhcb}, magenta: CMS~\cite{cms}) are in good agreement, ALICE is the only experiment at the LHC which is able to measure \jpsi{} down to $\pt{}=0$ and thus directly measure the \pt{} integrated cross section at midrapidity.
The right panel of Fig.~\ref{fig:jp_pp} shows the measured \pt{} spectra in the di-muon decay channel for the two different available beam energies.
A Next to Leading Order (NLO) NRQCD prediction \cite{butenschoen} is in good agreement with the data.
For the first time, the dependence of the \jpsi{} production on the charged particle multiplicity has been measured in pp collisions.
Figure \ref{fig:jpmulti} shows the results of the analyses in the \epem{} (red circles) and the \mpmm{} (blue triangles) decay channels.
The multiplicity is normalized by the mean charged particle multiplicity~\cite{multipaper} and has been determined in a pseudorapidity range of $|\eta|<1.0$ by counting the number of tracklets found in the ITS pixel detector.
The \jpsi{} yield per event in a given multiplicity bin is normalized to the inclusive yield per inelastic pp collision.
An approximately linear increase with multiplicity is observed in both rapidity ranges.
In the highest multiplicity bin, corresponding to approximately 5 times the minimum bias multiplicity, a \jpsi{} yield of around 6 times the minimum bias yield is found.
The strong dependence of the \jpsi{} production on the event multiplicity is an intriguing result that needs a theoretical interpretation.
\begin{figure}[tb]
  \begin{center}
    \includegraphics[width=0.5\textwidth]{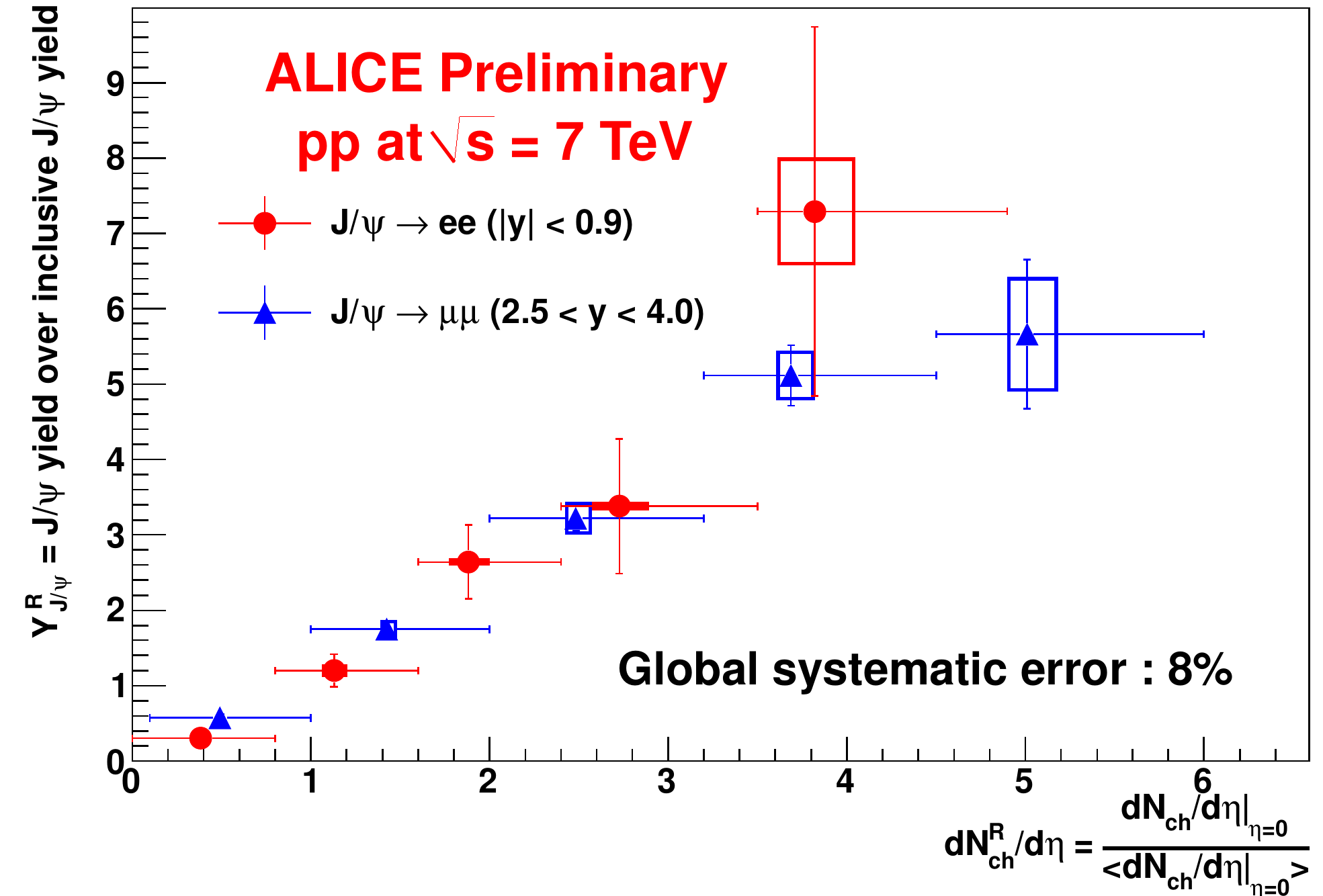}
    \caption{The measured multiplicity dependence of \jpsi{} production in 7 TeV pp collisions}
    \label{fig:jpmulti}
  \end{center}
\end{figure}
%
\begin{figure}[tb]
  \begin{center}
    \includegraphics[width=1.0\textwidth]{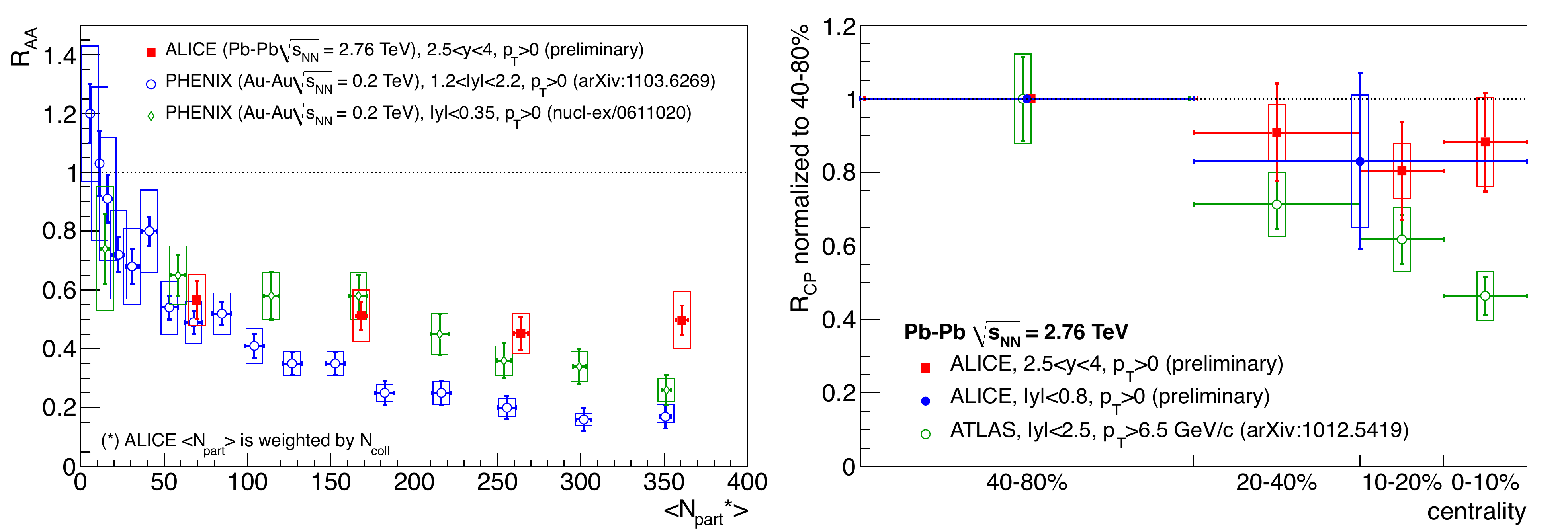}
    \caption{Left: \raa{} at $\rsnn{}=2.76$~TeV compared to PHENIX results~\cite{phenix} at $\rsnn{}=0.2$~TeV, right: \rcp{} at $\rsnn{}=2.76$~TeV compared to the ATLAS result~\cite{atlasRCP}}
    \label{fig:raarcp}
  \end{center}
\end{figure}
%

In fall 2010 the LHC collided Pb beams for the first time. 
Table~\ref{tab:data} summarizes the integrated luminosity used for the analysis.
The collision energy of $\rsnn{}=2.76$~TeV is almost 14 times higher than the energies reached at RHIC before.
Having pp data recorded at the same energy allows to calculate the nuclear modification factor $\raa{}=Y_{\rm{}Pb-Pb}/\left<N_{\rm{}coll}\right>Y_{\rm{}pp}$ for different Pb--Pb collision centralities.
$\left<N_{\rm{}coll}\right>$ corresponds to the average number of binary nucleon-nucleon collisions in a given centrality class.
A \raa{} equal to unity indicates no modification in heavy-ion collisions with respect to pp collisions.
The left panel of Fig.~\ref{fig:raarcp} shows as red squares the result of this measurement as a function of $\left<N_{\rm{}part}\right>$, the number of nucleons participating in the collision.
To account for the bias due to the large centrality bins, $\left<N_{\rm{}part}\right>$ has been weighted by $N_{\rm{}coll}$~\cite{pillot}.
Already for peripheral collisions ($\left<N_{\rm{}part}\right>\simeq{}70$), a suppression of the \jpsi{} yield is observed and \raa{} is about 0.6.
For more central collisions, the \raa{} exhibits only a weak change.
The data are compared to results from the PHENIX experiment at RHIC~\cite{phenix} in green diamonds at central and in blue circles at forward rapidity.
For central collisions the observed suppression is smaller than at RHIC, yet it has to be taken into account that shadowing and nuclear-absorption cross sections are expected to be different at the two energies.
Also, the energy density at a given $N_{\rm{}part}$ presumably differs.
A closer look into the difference of central and peripheral collisions can be obtained by calculating the central-to-peripheral modification factor $\rcp{}=\left<N_{\rm{}coll}^{\rm{}per.}\right>Y_{\rm{}Pb-Pb}^{\rm{}cent.}/\left<N_{\rm{}coll}^{\rm{}cent.}\right>Y_{\rm{}Pb-Pb}^{\rm{}per.}$.
There, the yields in the different centrality classes are divided by the corresponding average number of binary collisions.
As reference, the most peripheral bin is used.
Due to limitations in statistics it is currently set to a rather broad range of 40--80~\%.
An advantage of this variable is that many systematic uncertainties cancel out in the ratio.
The right panel of Fig.~\ref{fig:raarcp} shows the measured \rcp{} as a function of the centrality.
Red squares correspond to the ALICE result at forward rapidity.
The challenging measurement in the di-electron channel at midrapidity (blue circles) agrees within its errors; its precision will be improved in the next LHC Pb--Pb run.
A similar conclusion can be drawn as from the \raa{} result: compared to the most peripheral bin, a weak centrality dependence is observed.
On the other hand, high \pt{} \jpsi{}, measured by ATLAS above 6.5~GeV/$c$, show a stronger suppression in central (0--10~\%) events~\cite{atlasRCP}.

In summary, the inclusive \jpsi{} production has been measured both in pp and in Pb--Pb collisions.
In pp, differential cross sections have been obtained for two different beam energies and rapidity ranges.
An approximately linear increase of the \jpsi{} production with the charged particle multiplicity is observed both at central and forward rapidity, which calls for theoretical interpretation.
In Pb--Pb, both nuclear modification factors \raa{} and \rcp{} have been measured.
A significant suppression of the \jpsi{} yield is observed, with a weak dependence on centrality.
At central events, a smaller suppression has been found compared to PHENIX results at much lower collision energies.



%

}  
